\documentclass[a4paper,11pt]{article}
\usepackage{latexsym}
\usepackage{epsfig, subfigure, subfloat, graphicx, float }
\usepackage{amssymb}
\usepackage{amsmath}
\usepackage{epstopdf}
\usepackage{authblk}
\usepackage{cite}
\author{H. Mohseni Sadjadi\footnote{mohsenisad@ut.ac.ir}  and V. Anari\footnote{v.anari@ut.ac.ir}
\\ {\small Department of Physics, University of Tehran,}
\\ {\small P. O. B. 14395-547, Tehran 14399-55961, Iran}}
\title{Mass varying neutrinos, symmetry breaking, and cosmic acceleration}

\begin{document}
\maketitle
\begin{abstract}
We introduce a new proposal for the onset of cosmic acceleration based on mass-varying neutrinos. When massive neutrinos become nonrelativistic, the $Z_2$ symmetry breaks, and the quintessence potential becomes positive from its initially zero value. This positive potential behaves like a cosmological constant at the present era and drives the Universe's acceleration during the slow roll evolution of the quintessence. In contrast to $\Lambda$CDM model, the dark energy in our model is dynamical, and the acceleration is not persistent. Contrary to some of the previous models of dark energy with mass-varying neutrinos, we do not use the adiabaticity condition which leads to instability.

\end{abstract}

\section{Introduction}
The origin of the present acceleration of the Universe \cite{acc1,acc2,acc3,acc4,acc5,acc6,acc7}is not yet known. One can attribute this acceleration to an exotic matter with negative pressure that permeates the Universe homogeneously, dubbed dark energy.  A simple candidate for dark energy is a scalar field known as quintessence, which constitutes nearly 70\%  of our present Universe with density $\rho_{dark}\sim 10^{-10}eV^4$ \cite{quint01,quint02,quint1,quint2,quint3,quint4,quint5,quint6,quint7,quint8,quint9,quint10,quint11,quint12,quint13,quint14}. Structure formation requires that the acceleration begin after the matter-dominated era. The equation of state (EoS) parameter of the quintessence is negative, and it dilutes less quickly than dark and ordinary matter and radiation. Today, dark energy density has the same order of magnitude as the sum of other Universe ingredients, hence in the early eras, its ratio density was negligible. "Why, nowadays, dark energy and dark matter densities have the same order of magnitude?"  is known as the coincidence problem \cite{coinc1,coinc2,coinc3,coinc4,coinc5,coinc6,coinc7,coinc8,coinc9,coinc10}. This can be reexpressed as why in the early time, the dark energy density was negligible.  The present proportion of dark sectors can be explained by considering the possible interaction of dark energy  with other components \cite{inter1,inter01,inter2,inter3,inter4,inter5,inter6,inter7,inter8,inter9,inter10,inter11,inter12,inter13,inter14,inter15,inter16,inter17}. Known physical properties of these components may give some clues to understand the behavior of dark energy.
For example, quintessence (also dubbed the acceleron in neutrino dark energy models) and neutrinos interaction,  may be employed to relate neutrino masses to EoS and the density of dark energy \cite{far1,far2,far3}. This interaction makes the mass of neutrino a function of the quintessence. Hence the neutrino mass changes by the evolution of the scalar field.
The transition of mass varying neutrinos from relativistic to non-relativistic phase deforms the effective potential such that the quintessence velocity decreases, and it follows the minimum of the convex effective potential, gives rise to the Universe acceleration \cite{far1,far2,far3,far4,far5,far6,far7,far8,far9}. In some papers, an adiabatic evolution for the quintessence is considered \cite{far1,afsh}, such that the quintessence effective mass becomes larger than the Hubble parameter. This scenario may suffer from instabilities which result in the formation of neutrino nuggets \cite{afsh},\cite{mota}. These instabilities and the possibility to have stable neutrino lumps are also discussed in \cite{Wet1}, where lumps are considered as non-relativistic particles with effective interactions, and also in \cite{Wet2} for a large neutrino mass.

In another class of models \cite{sym1,sym2,sym3}, to describe the screening effect, a coupling between the quintessence and pressureless matter is considered. When the density of matter is greater than a critical value, the quintessence vacuum expectation value vanishes, leading to zero fifth force. But when the matter density becomes less than the critical value (e.g. by the redshift), the $Z_2$ symmetry is broken, and the quintessence evolves towards the minimum of its effective potential.  This evolution may describe the present acceleration of the Universe. But in the symmetron model, the quintessence is too heavy to slow roll, and instead, rolls rapidly toward the minimum of its effective potential and oscillates about it. To remedy this problem, in \cite{sad1,sad2}, the symmetron is considered in the teleparallel model of gravity which has a de Sitter attractor solution in the late time.

In this article, we try to introduce a new model to explain the onset of the positive acceleration of the Universe from the matter dominated era with zero dark energy density. Motivated by the mass varying neutrino and the symmetron, we introduce a coupled quintessence neutrino model in which the potential and the neutrino mass have $Z_2$ symmetry. By the evolution of mass varying neutrinos from the relativistic regime to the nonrelativistic one, the shape of the effective potential changes and the quintessence begins its evolution from a constant initial fixed point. This procedure may provide enough {\it{positive}} potential to drive the cosmic acceleration via a slow roll evolution from a decelerated epoch.

In our model the rise of dark energy and its dominance over other components depend on the neutrino mass which determines the time when the neutrinos become nonrelativistic. So the evolution of the quintessence from a zero density is postponed until the nonrelativistic era of neutrinos after which the equivalence of dark matter and dark energy densities may occur. In this way, one may relate the coincidence problem to the neutrino mass. The coincidence problem also depends on the other parameters of the model, especially those which determine the dark energy density.

As the adiabaticity condition (i.e. the quintessence adiabatically traces the minimum of the effective potential) is not used, the model is free from instabilities encountered in some of the growing neutrino quintessence models \cite{afsh}. Besides, in contrast to the symmetron model \cite{sym1}, the Universe can experience an accelerated phase during a time greater than the Hubble time in the slow roll regime.

The scheme of the paper is as follows:
In the second section, we study the possibility of the occurrence of cosmic acceleration triggered by massive neutrinos in a symmetronlike model, from an epoch with zero dark energy density. In the third section, the perturbation equations are obtained, and the stability of the model is discussed. We illustrate our results with some numerical examples. In the last section, we conclude the paper.

Throughout this paper we use units $\hbar=c=k_B=1$ and  metric signature (-,+,+,+).

\section{Cosmic acceleration triggered by massive neutrinos in quintessence models with $Z_2$ symmetry}
We use the action \cite{fl}
\begin{equation}\label{1}
S=\int d^4x\sqrt{-g}\left[{1\over 2}M_P^2R-{1\over 2}\partial_\mu \phi\partial^\mu \phi-V(\phi)\right]+\sum_jS_j\left[A_j^2(\phi)g_{\mu \nu},\psi_j\right]
\end{equation}
where $\phi(t)$ is the homogenous quintessence with potential $V(\phi)$, and $\psi_j$ denotes other species . The coupling between the quintessence and $\psi_j$ is given by the conformal coupling $A_j^2(\phi)g_{\mu \nu}$, where $A_j(\phi)>0$. We consider only an interaction between $\psi_{i}$ and the quintessence, then $A_{(j)}(\phi)=\delta_{ij}A(\phi)$. $M_P=2.4\times 10^{18}GeV$ is the reduced Planck mass. The Universe is taken as a spatially flat Friedmann-Lemaitre-Robertson-Walker (FLRW) space time
\begin{equation}\label{1.1}
ds^2=-dt^2+a^2(t)(dx^2+dy^2+dz^2),
\end{equation}
where $a(t)$ is the scale factor.

Variation of (\ref{1}) with respect to $\phi$ gives
\begin{equation}\label{2}
\ddot{\phi}+3H\dot{\phi}+V_{,\phi}= -{A_{,\phi}\over A}(\rho_{(i)}-3P_{(i)}),
\end{equation}
where $\rho_{(i)}$ and $P_{(i)}$ are the energy density and the pressure of the {\it{i}}th species respectively and $V_{,\phi}={dV\over d\phi}$.
Variation of (\ref{1}) with respect to the metric yields the Friedmann equation
\begin{equation}\label{3}
H^2={1\over 3M_P^2}\left({1\over 2}\dot{\phi}^2+V+\sum_j \rho_{(j)}\right),
\end{equation}
and the evolution of the Hubble parameter is given by
\begin{equation}\label{4}
\dot{H}=-{1\over 2M_P^2}\left(\dot{\phi}^2+\sum_j (\rho_{(j)} +P_{(j)})\right).
\end{equation}
The Universe is positively accelerated provided that, $\dot{H}+H^2>0$, which yields
\begin{equation}\label{14}
2V(\phi)-2\dot{\phi}^2-\sum_i(\rho_{(i)}+3P_{(i)})>0.
\end{equation}
The continuity equations are given by
\begin{equation} \label{5}
\dot{\rho}_{(i)}+3H(P_{(i)}+\rho_{(i)})={A_{,\phi}\over A}\dot{\phi}(\rho_{(i)}-3P_{(i)}),
\end{equation}
for interacting {\it{i}}th species, and
\begin{equation} \label{6}
\dot{\rho}_{(j)}+3H(P_{(j)}+\rho_{(j)})=0,
\end{equation}
for other components. The neutrino-quintessence interaction resulting from (\ref{1}) can also be considered in the context of the coupled quintessence model \cite{quint01},\cite{inter1},\cite{inter01}.

By employing the Fermi-Dirac distribution for neutrinos whose masses $m_{(\nu)}(\phi)$ are $\phi$ dependent and are also in thermal equilibrium with temperature $T_{(\nu)}$,  one obtains
\begin{eqnarray}\label{fd}
&&\rho_{(\nu)}={T_{(\nu)}^4\over \pi^2}\int_0^\infty {dx x^2 \sqrt{x^2+\xi^2}\over e^x+1}\nonumber \\
&&P_{(\nu)}={T_{(\nu)}^4\over 3\pi^2}\int_0^\infty {dx x^4\over \sqrt{x^2+\xi^2}(e^x+1)},
\end{eqnarray}
where $\xi={m_{(\nu)}(\phi)\over T_{(\nu)}}$. By using  (\ref{fd}) one finds
\begin{equation}\label{7}
\dot{\rho}_{(\nu)}+3H(P_{(\nu)}+\rho_{(\nu)})={m_{(\nu),\phi}(\phi)\over m_{(\nu)}(\phi)}\dot{\phi}(\rho_{(\nu)}-3P_{(\nu)}).
\end{equation}
Therefore (\ref{5}) is the same as the mass varying neutrino continuity equation provided that $A(\phi)={m_{(\nu)}(\phi)\over M}$, where $M$ is a mass scale. For the quintessence we have
\begin{equation}\label{7.1}
\ddot{\phi}+3H\dot{\phi}+V_{eff.,\phi}= 0,
\end{equation}
where the effective potential is given by
\begin{equation}\label{7.2}
V_{eff.,\phi}=V_{,\phi}+{m_{(\nu),\phi}(\phi)\over m_{(\nu)}(\phi)}(\rho_{(\nu)}-3P_{(\nu)}).
\end{equation}
So we take the neutrinos interacting with quintessence via (\ref{1}) as mass varying neutrinos.  For different kinds of neutrinos with a same ${m_{(\nu),\phi}(\phi)\over m_{(\nu)}(\phi)}$, we may still use (\ref{7.2}) and (\ref{7}), provided that we take $\rho_{(\nu)}=\sum_i  \rho_{(\nu_i)}$ and
$P_{(\nu)}=\sum_i  P_{(\nu_i)}$ .

When interacting neutrinos are relativistic, $m_{(\nu)}\ll T_{(\nu)}$,  we have
\begin{eqnarray}\label{8}
\ddot{\phi}+3H\dot{\phi}+V_{,\phi}= 0 \nonumber \\
\dot{\rho}_{(\nu)}+4H\rho_{(\nu)}=0,
\end{eqnarray}
and $V_{eff.}=V$,
while for nonrelativistic ones, $m_{(\nu)}\gg T_{(\nu)}$, we have
\begin{eqnarray}\label{9}
\ddot{\phi}+3H\dot{\phi}+V_{,\phi}= -{m_{(\nu),\phi}\over m_{(\nu)}}\rho_{(\nu)} \nonumber \\
\dot{\rho}_{(\nu)}+3H\rho_{(\nu)}={m_{(\nu),\phi}\over m_{(\nu)}}\dot{\phi}\rho_{(\nu)}.
\end{eqnarray}
We can define a rescaled energy density $\hat{\rho}_{(\nu)}$  as
\begin{equation}\label{10}
\rho_{(\nu)}=m_{(\nu)} \hat{\rho}_{(\nu)},
\end{equation}
in terms of which (\ref{9}) reduces to
\begin{eqnarray}\label{11}
\ddot{\phi}+3H\dot{\phi}+V_{,\phi}+m_{(\nu),\phi}\hat{\rho}_{(\nu)}=0 \nonumber \\
\dot{\hat{\rho}}_{(\nu)}+3H{\hat{\rho}}_{(\nu)}=0.
\end{eqnarray}
In the nonrelativistic case , we can write  $\rho_{(\nu)}=m_{(\nu)} n_{(\nu)}$, where $n_{(\nu)}$ is the neutrino density number. So we identify $n_{(\nu)}=\hat{\rho}_{(\nu)}$. The solution of the second equation in (\ref{11}) is
\begin{equation}\label{12}
\hat{\rho}_{(\nu)}=\hat{\rho}_{(\nu)}^0a^{-3},
\end{equation}
where "0" denotes the present time, where we take $a_0=1$. Equivalently  $n_{(\nu)}=n_{(\nu)}^0a^{-3}$, as expected. From (\ref{11}), in the nonrelativistic limit we can define an effective quintessence potential
\begin{equation}\label{13}
V_{eff.,\phi}=V_{,\phi}+m_{(\nu),\phi}n_{(\nu)}.
\end{equation}

Now we can construct our model. We require

i) Initially, when massive neutrinos are relativistic, quintessence energy density be negligible, and the Universe be in a decelerated phase.

ii) The accelerated expansion of the Universe be caused by symmetry breaking triggered by the evolution of mass varying neutrinos from the relativistic regime toward the nonrelativistic one.

 To choose appropriate $V(\phi)$ and $m_\nu(\phi)$  to fulfil (i) and (ii), we proceed as follows:

  We assume that $V(\phi)$ and $m(\phi)$ have $Z_2$ symmetry and initially the quintessence stays at the minimum of its potential which we take : $V_{min.}=V(\phi*)=0$. Thus dark energy density is negligible in this era
  \begin{equation}
  \rho_\phi={1\over 2}\dot{\phi}^2+V(\phi)=V(\phi*)=0.
  \end{equation}
  To have an initial stable solution, we require that the potential be convex at this point $V_{,\phi \phi}{(\phi=\phi*)}>0$.   As neutrinos are initially relativistic: $\rho_{(\nu)}\approx 3P_{(\nu)}$, we have $V=V_{eff.}$. From (\ref{14}) we find that the Universe is in a decelerated phase. In this era as $\phi$ is a constant, neutrino masses are also constant and the interaction in (\ref{7}) is nonoperative. Due to the Universe's expansion, neutrinos exit from relativistic phase such that $(\rho_{(\nu)}-3P_{(\nu)})$ becomes significant,  $\rho_{(\nu)}-3P_{(\nu)}>0$. Hence the effective potential, given by (\ref{13}), is no longer equal to the quintessence potential. If we choose $m_{(\nu),\phi\phi}{(\phi=\phi*)}<0$, whenever $(\rho_{(\nu)}-3P_{(\nu)})>-{V_{,\phi \phi}{(\phi=\phi*)}\over m_{(\nu),\phi\phi}{(\phi=\phi*)}}$ , the effective potential becomes concave and $\phi*$ becomes an unstable point. Therefore the quintessence rolls down the effective potential and the $Z_2$ symmetry breaks. Contrary to the effective potential, the potential is convex, and the quintessence climbs its own potential. This can be achieved only when $V_{eff.,\phi}$ and $V_{,\phi}$ have opposite signs. From (\ref{7.2}) this implies that the signs of $m_{(\nu),\phi}$  and $V_{,\phi}$  are opposite too. This mechanism provides the positive potential required for cosmic acceleration (see (\ref{14})).

This scenario is entirely different from the usual growing neutrino quintessence studied in the literature. In that scenario the interaction of neutrinos and quintessence, after the neutrinos become nonrelativistic, acts as a barrier potential and stops the fast rolling of the quintessence forcing it to follow the minimum of the effective potential giving rise to cosmic acceleration.  In some papers an adiabatic evolution for the quintessence is considered \cite{far1,afsh}. This adiabaticity, which is absent in our model, gives rise to neutrino perturbation growing and neutrino nugget formation \cite{afsh}.  Our model is also different from the symmetron model where $V_{eff.,\phi}=V_{,\phi}+A_{,\phi}\hat{\rho}$, and $\hat{\rho}$ is the rescaled pressureless matter density.  In the symmetron model, by the dilution of matter density, the quintessence becomes tachyonic and rolls down simultaneously its own and its effective potential \cite{dark}. Therefore the potential decreases by the symmetry breaking, and if it is initially negligible, it will become negative after a while and cannot drive the acceleration \cite{dark}. So in the symmetron model, the initial dark energy density is assumed to be non-negligible $\rho_{\phi*}=\Lambda>0$.

Based on astrophysical data the EoS of the quintessence,
\begin{equation}
w_\phi ={{1\over 2}\dot{\phi}^2-V\over {1\over 2}\dot{\phi}^2 +V},
\end{equation}
is estimated to be  $w_\phi =-1.006 \pm 0.045$ in the present epoch\cite{Planck} . So the kinetic energy of the quintessence must be much less than its potential. This is the slow roll condition
\begin{equation}\label{18}
{1\over 2}\dot{\phi}^2\ll V(\phi).
\end{equation}
From (\ref{11}) we have
\begin{equation}\label{19}
\dot{\phi}=-{V_{eff.,\phi}\over 3H(1+\chi)},
\end{equation}
where $\chi={\ddot{\phi}\over 3H\dot{\phi}}$.  The slow roll condition is satisfied when
\begin{equation}\label{20}
{1\over 2}\left({V_{eff.,\phi} \over 3H(1+\chi)}\right)^2\ll V.
\end{equation}
If $\chi\sim \mathcal{O}(1)$ or $\chi \lesssim 1$ , (\ref{20}) becomes
\begin{equation}\label{21}
V_{eff.,\phi}^2\ll 9H^2 V.
\end{equation}
If in the slow roll epoch (like the present era), dark energy and other components densities have the same order of magnitude, by $3M_P^2H^2\sim V$  and (\ref{21}) we obtain
\begin{equation}\label{22}
V_{eff.,\phi}\ll {3V^2\over M_P^2}.
\end{equation}

As a summary, in our formalism of cosmic acceleration,  when the mass varying neutrinos become nonrelativistic their interaction with the quintessence becomes operative and triggers the quintessence evolution which augments the potential from its initial zero value. This positive potential is necessary to drive the cosmic acceleration. This mechanism provides a slow roll evolution provided that the effective potential is sufficiently flat (in the sense used in (\ref{22})).

To get more intuition about our model let us give an example. We choose the potential as a combination of cosmological constant and a Gaussian type potential \cite{pot1,pot2}. We assume that the mass of neutrinos also has a Gaussian form \cite{mass1,mass2}.
\begin{eqnarray}\label{14.1}
V(\phi)&=&V_0(1-e^{-\alpha \phi^2})\nonumber \\
m_{(\nu)}(\phi)&=&m^*e^{-\beta \phi^2}
\end{eqnarray}
where $\alpha>0$ and $\beta>0$ are constants with inverse mass squared dimensions and $V_0>0$.  $m^*$ is the neutrino mass at $\phi=0$.

Initially, neutrinos are relativistic and $V_{eff.}=V$.  The quintessence effective mass is assumed to be much less than the Hubble parameter,  $2V_0\alpha \ll H^2$, such that (\ref{11}) describes an overdamped oscillation equation \cite{sym2}.  Therefore, in this epoch $\phi=0$ is a stable solution (as the potential is convex) of equations of motion, yielding a negligible dark energy density  $\rho_\phi(\phi=0)=0$.
When temperature decreases,   $\rho_{(\nu)}-3P_{(\nu)}$  is no longer negligible and
\begin{equation}\label{14.2}
V_{eff.,\phi}=2\alpha V_0 \phi e^{-\alpha \phi^2}-2\beta\phi (\rho_{(\nu)}-3P_{(\nu)}).
\end{equation}
When ${\alpha V_0\over \beta}<\rho_{(\nu)}-3P_{(\nu)}$, the effective potential becomes concave at $\phi=0$ and this point becomes unstable and the quintessence, which gains a negative mass squared, rolls down the effective potential while climbing its own potential (since $V_{eff.,\phi}$ and $V_{,\phi}$ have different signs). This holds whenever
\begin{equation}\label{14.31}
{\alpha \over \beta} V_0<  e^{\alpha \phi^2}(\rho_{(\nu)}-3P_{(\nu)}).
\end{equation}
This mechanism provides the positive potential needed for the acceleration.

In the nonrelativistic limit $m^*e^{-\beta \phi^2}\gg T_{(\nu)}$, we can ignore the pressure. The effective potential becomes
\begin{equation}\label{15}
V_{eff.}=V_0(1-e^{-\alpha \phi^2})+n_{(\nu)} m^*e^{-\beta \phi^2},
\end{equation}
and we can write (\ref{14.31}) as
\begin{equation}\label{14.32}
V_0< {\beta\over \alpha}m^*n_{(\nu)} e^{(\alpha-\beta) \phi^2}.
\end{equation}

\begin{figure}[H]
\centering\epsfig{file=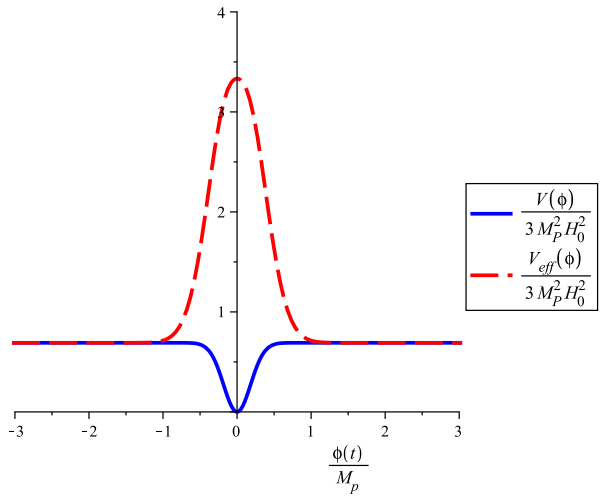,width=8cm,angle=0}
\caption{ The potential and the effective potential for $\{\alpha=15M_P^{-2},\,\,\,\beta=15M_P^{-2},\,\,\,\, V_0=0.691 \times 3H_0^2 M_P^2,\,\,\,\rho_{(\nu)}=10 H_0^2M_P^2\}$. } \label{fig1}
\end{figure}

 When $\phi^2$ increases, $e^{-\alpha \phi^2}\ll 1$ holds and the potential behaves as a cosmological constant (see Fig.(\ref{fig1})) at late time. We take this era as our present era and as based on astrophysical data, $\rho_\phi$ constitutes about 0.7 of our Universe density, we can take $V_0\sim \left({7\over 10}\right)3M_P^2 H_0^2$.  Because of the exponential factors in (\ref{14.1}), the derivative of the effective potential satisfies (\ref{21}) when $\alpha\phi^2\sim 1$ and $\beta\phi^2\sim 1$, implying a slow roll motion with $w_{\phi}\simeq -1$.
 Eventually, by dilution of massive neutrinos, the effective potential becomes the same as the potential and the quintessence rolls down towards its initial point and oscillates around it.

Obtaining analytic solutions for the equations of motions, even with simple potential and mass function,  is very complicated if not impossible. So let us illustrate our results via a numerical example by using eqs.(\ref{3}),(\ref{4}),(\ref{6}), and (\ref{11}). We assume that the Universe is nearly composed of massive neutrinos $(\nu)$,  the quintessence $(\phi)$, the pressureless matter $(c)$ comprising cold dark matter and pressureless baryonic matter, and radiation $(r)$. We choose the parameters of the model and the initial conditions as $\{\alpha=15M_P^{-2},\,\,\,\beta=15M_P^{-2},\,\,\,\, V_0=0.691 \times 3H_0^2 M_P^2=2.74\times 10^{-47} GeV^4 \}$ and
\begin{eqnarray} \label{IC1}
&&{\phi\over M_P}=10^{-10},\,\, \dot{\phi}=10^{-6}M_P H_0,\,\,\,\rho_{(\nu)}= 3.4\times 10^{10}H_0^2M_P^2,\nonumber \\
&& \rho_{(c)}=1.54\times 10^{11} H_0^2 M_P^2,\,\,\, \rho_{(r)}=2.50\times 10^{11}H_0^2M_P^2.
\end{eqnarray}
respectively. The initial conditions are set at
$\tau=tH_0=0$ which in our model is equivalent to the redshift $z=5500$ corresponding to the radiation-dominated Universe. $H_0$ is the present Hubble parameter, i.e. the Hubble parameter at $a=1$.
The relative densities defined by $\Omega_{(i)}={\rho_{(i)}\over 3 M_P^2H^2}$ are derived from (\ref{IC1}) as
\begin{equation} \label{IC11}
\Omega_{(r)}=0.571,\,\,\, \Omega_{(\nu)}=0.077,\,\, \Omega_{(c)}=0.352,\,\,\Omega_{(\phi)}=1.14\times 10^{-24},
\end{equation}
and the Hubble parameter is $H=3.82\times 10^5 H_0$. $\Omega_{(\phi)}=1.14\times 10^{-24}$ shows that the initial values chosen for $\phi$ and $\dot{\phi}$ give only a negligible dark energy contribution in the total density. In our numerical study we assume that neutrinos are completely nonrelativistic at $\tau=0$, i.e.  $\rho_{(\nu)}-3P_{(\nu)}\simeq \rho_{(\nu)}$. So we can ignore the neutrinos pressure.  In order that $\rho_{(\nu)}-3P_{(\nu)}\simeq \rho_{(\nu)}$ holds, we must have
$m^{*}\gg T_{(\nu)}$ at $\tau=0$. The mass varying neutrinos exit from the relativistic regime when $m^*\simeq 3T_{(\nu)}^*$ corresponding to the redshift $z=z_{nr}$. Until this time we have \cite{Liddle}
\begin{equation}\label{T1}
T_{(\nu)}=\left({4\over 11}\right)^{1\over 3}T_{\gamma},
\end{equation}
where $T_{\gamma}$ is the photons temperature. In addition we have \cite{Liddle}
\begin{equation}\label{T2}
T_{\gamma}=T_{\gamma}^0(1+z),
\end{equation}
where $T_{\gamma}^0$ is the photons temperature at the present time. Hence
\begin{equation}\label{T3}
T_{(\nu)}^*=\left({4\over 11}\right)^{1\over 3}T_{\gamma}^0(1+z_{nr})=0.085\times 10^{-3}(1+z_{nr}).
\end{equation}
 Therefore in our example we must have $m^*\gg 0.92 eV$.

 It is worth Note that we have chosen our initial condition at $\tau=0$ in the nonrelativistic regime while the quintessence began its motion in the semirelativistic regime, where $P_{(\nu)}$ was not negligible, therefore the values in (\ref{IC1}) are not the variables values just after the symmetry breaking. Our numerical results illustrate the evolution of the Universe from an epoch with $\Omega_{(\phi)}\simeq 0$, to the present dark-energy-dominated epoch. We also study the future behavior of the quintessence. A quantitative study beginning from the semirelativistic regime of neutrinos requires considering the pressure $P_{(\nu)}$ (see (\ref{fd})), which makes the equations very complicated to solve. The initial conditions for the scalar field are due to the quantum fluctuations around $\phi=0$, against which the model is no more stable after the symmetry breaking. Therefore by a small deviation from $\phi=0$, the quintessence rolls down its steep effective potential \cite{sym1}.

In Fig.(\ref{fig1}), we have depicted the potential and the effective potential for $\rho_{(\nu)}=10 H_0^2M_P^2$.  The potential is the same as the effective one in the relativistic limit. For nonrelativistic massing neutrinos the shape of the effective potential changes, and the previous minimum point becomes the new maximum.

\begin{figure}[H]
\centering\epsfig{file=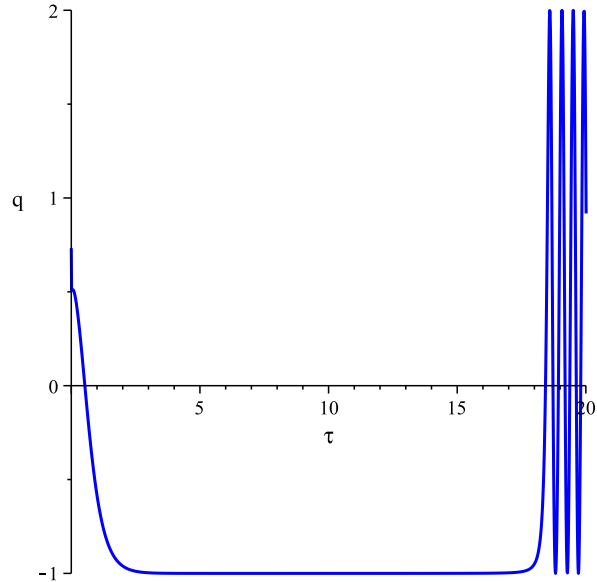,width=8cm,angle=0}
\caption{ The deceleration parameter in terms of dimensionless time $\tau=tH_0$, for $\{\alpha=15M_P^{-2},\,\,\,\beta=15M_P^{-2},\,\,\,\, V_0=0.691 \times 3H_0^2 M_P^2\}$ and
initial conditions (\ref{IC1}) } \label{fig2}.
\end{figure}

In Fig.(\ref{fig2}), the deceleration parameter,  $q=-{\ddot{a}a\over \dot{a}^2}=-\left(1+{\dot{H}\over H^2}\right)$, is depicted showing the transition of the Universe from a deceleration epoch to acceleration in a time of order of the Hubble time.

\begin{figure}[H]
\centering\epsfig{file=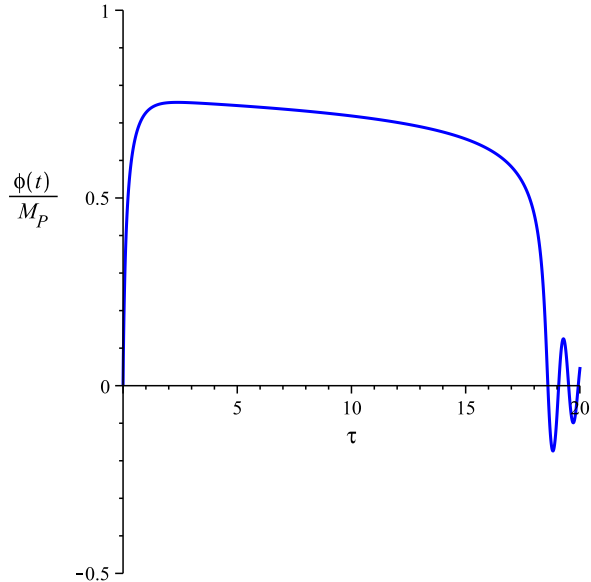,width=8cm,angle=0}
\caption{ The quintessence field in terms of dimensionless time $\tau=tH_0$, for $\{\alpha=15M_P^{-2},\,\,\,\beta=15M_P^{-2},\,\,\,\, V_0=0.691 \times 3H_0^2 M_P^2\}$ and
initial conditions (\ref{IC1}) }.  \label{fig3}
\end{figure}

Although this acceleration, which begins at the redshift $z\simeq0.6$, can last for some Hubble times but is not persistent. Gradually as the neutrinos density dilutes, the effective potential becomes the same as the potential and the quintessence rolls down to its initial position and oscillates about it via an underdamped oscillation. This is due to the fact that   the quintessence effective mass becomes larger than the Hubble parameter at late time. This can also be seen from  Fig.(\ref{fig3}).  Fig.(\ref{fig3}) shows that the quintessence grows from $\phi=0$ and reaches to an approximately constant value, which is consistent with our previous discussion that when the effective potential becomes nearly flat, the slow roll evolution begins.

Before the symmetry breaking, (\ref{7.1}) is an overdamped harmonic oscillator equation, and $\phi=0$ is a stable point. After the symmetry breaking, at $\tau<0$, this point becomes unstable against fluctuations and $\phi$ commences its evolution. So just after the symmetry breaking, $\phi<\phi(\tau=0)$. Note that
$V(\phi(\tau=0))\ll V_0$, so $\phi$  cannot overcome the potential initially, and needs a time of the order of the present Hubble time to reach the maximum of its potential to drive the cosmic acceleration. In the future, by dilution of neutrino the quintessence will come back to its initial position, but because the Hubble parameter will be much smaller than the effective mass, the quintessence will have an underdamped oscillation (see Fig.(\ref{fig3})).

The effective potential becomes very steep after the symmetry breaking (see Fig.(\ref{fig1})), so we expect that $\dot{\phi}$ increases initially.  In our example $V_{eff.,\phi}\simeq -2\beta \rho_{(\nu)}\phi\sim -100H_0^2M_P$ which is more efficient than the friction term $3H\dot{\phi}\sim M_PH_0^2$. By the increase of $\dot{\phi}$ and decrease of $\rho_{(\nu)}$, the friction term becomes more relevant providing the required condition for the slow roll (see Fig.(\ref{fig3})).

\begin{figure}[H]
\centering\epsfig{file=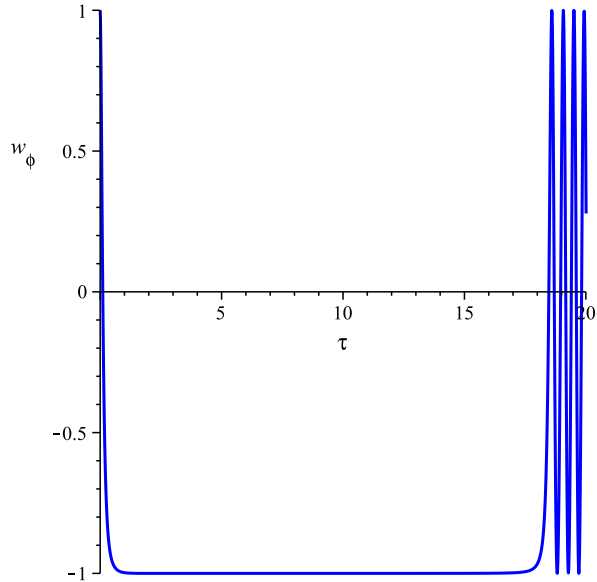,width=8cm,angle=0}
\caption{ The quintessence equation of state parameter in terms of dimensionless time $\tau=tH_0$, for $\{\alpha=15M_P^{-2},\,\,\,\beta=15M_P^{-2},\,\,\,\, V_0=0.691 \times 3H_0^2 M_P^2\}$ and
initial conditions (\ref{IC1}) } \label{fig4}
\end{figure}

The EoS parameter of the quintessence, $w_{\phi}$, is plotted in Fig.(\ref{fig4}), showing that $w_\phi$ decreases and $w_\phi\approx -1$ during the time where the effective potential is nearly flat. Finally in the future due to the quintessence oscillation, $w_{\phi}$ will oscillate between $-1$ and $1$.
In the present era $\tau=0.94$ (corresponding to $a=1$) we find $w_\phi=-0.998$ which is the range expected from Planck 2015 data.

\begin{figure}[H]
\centering\epsfig{file=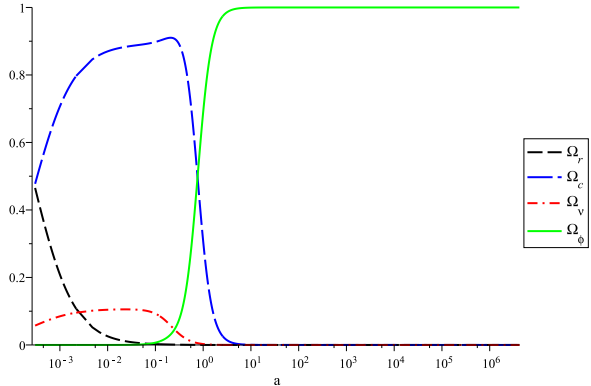,width=8cm,angle=0}
\caption{ Relative densities in terms of the scale factor $a$, for $\{\alpha=15M_P^{-2},\,\,\,\beta=15M_P^{-2},\,\,\,\, V_0=0.691 \times 3H_0^2 M_P^2\}$ and
initial conditions (\ref{IC1}) }.  \label{fig5}
\end{figure}

 The relative densities defined by $\Omega_{(i)}={\rho_{(i)}\over 3 M_P^2H^2}$, are depicted in Fig.(\ref{fig5}), showing that the  dark energy density grows while other ingredients ratio densities decrease.  $\Omega_{(c)}$, $\Omega_{(\nu)}$, $\Omega_{(r)}$, and $\Omega_{(\phi)}$ are the relative densities of the pressureless matter, the mass varying neutrinos, the radiation, and the dark energy respectively.

In this example, relative densities at the present era $\tau=0.94$ (corresponding to $a=1$) are obtained as
$\Omega_{(\phi)}=0.691,\,\,\, \Omega_{(c)}=0.308,\,\,\, \Omega_{(\nu)}=0.00003,\,\,\, \Omega_{(r)}=0.00009$ which lie in the expected region estimated by Planck 2015 data.

\section{Linear Perturbations}
In this section we consider the evolution equations of perturbations in the nonrelativistic era, $m_{(\nu)}\gg T_{(\nu)}$. We study the neutrino contrast with the same method used in \cite{mota}. In the mass varying models of dark energy based on adiabaticity, the linear perturbations grow and give rise to instability and the formation of neutrino nuggets \cite{afsh}. We first gather the required equations corresponding to our problem, which are derived in \cite{pert1,pert2,pert3}.  Then based on these equations, we will continue our discussion through a numerical illustrative example.

The line element of perturbed FLRW space-time can be written as
\begin{equation}\label{p1}
ds^2=-(1+2\varphi)dt^2+2a(t)B_{,i}dtdx^i+a^2(t)(\delta_{ij}+2(E_{,ij}-\psi \delta_{ij}))dx^idx^j,
\end{equation}
where $\varphi$ (lapse function), $B$ (shift function), $E$, and $\psi$ are four scalar functions and a comma denotes a partial derivative. The stress tensor perturbations are given by
\begin{eqnarray}\label{p2}
&&\delta T_{00}=\sum_j \delta \rho_{(j)}-\varphi\dot{\bar{\phi}}^2+\delta \dot{\phi}\dot{\bar{\phi}}+V'(\bar{\phi})\delta \phi \nonumber \\
&&\delta T_{0i}=a\left(\dot{\bar{\phi}}(\dot{\bar{\phi}}B_{,i}+{1\over a}\delta\phi_{,i})-\sum_j(\bar{\rho}_{(j)}+\bar{P}_{(j)})v_{(j),i} \right)\nonumber \\
&&\delta T_{ij}=\delta_{ij}a^2\left( \sum_j\delta P_{(j)}-\varphi\dot{\bar{\phi}}^2+\delta \dot{\phi}\dot{\bar{\phi}}-V'(\bar{\phi})\delta \phi\right).
\end{eqnarray}
 By bar we denote the background value of a parameter and a prime denotes derivative with respect to the argument. Four velocities of the fluids are given by
 \begin{eqnarray}\label{p3}
 u_{(j)0}=-(1+\varphi)\nonumber \\
 u_{(j)i}=a(v_{(j)}+B)_{,i}
 \end{eqnarray}
where $\bar{u}_{(j)0}=-1$ and $\bar{u}_{(j)i}=0$ have been used.  Going to the Fourier space, the evolution equations for density fluctuations are derived as
\begin{eqnarray}\label{p4}
&&\delta \dot{\rho}_{(j)}-\left({k^2 v_{(j)}\over a}+k^2E+3\dot{\psi}\right)(\bar{\rho}_{(j)}+\bar{P}_{(j)})+3H(\delta \rho_{(i)}+\delta P_{(j)})=\nonumber \\
&&\beta_{(j)}(\phi)(\bar{\rho}_{(j)}-3\bar{P}_{(j)})\delta \dot{\phi}+\beta_{(j)}(\phi)(\delta \rho_{(i)}-3\delta P_{(j)})\dot{\bar{\phi}}+\nonumber \\
&&\beta'_{(j)}(\phi)(\bar{\rho}_{(j)}-3\bar{P}_{(j)})\dot{\bar{\phi}}\delta \phi,
\end{eqnarray}
in which $\beta_{(\nu)}={m'_{(\nu)}(\phi)\over  m_{(\nu)}(\phi)}$ and $\beta_{(j)}=0$ for $(j)\neq (\nu)$, this means that the interaction is considered only between the quintessence and massive neutrinos.
From momentum conservation, the constraint
\begin{eqnarray}\label{p5}
&&\dot{v}_{(j)}=-{\beta_{(j)}(\bar{\phi})\over a}{\bar{\rho}_{(\nu)}-3\bar{P}_{(\nu)}\over \bar{\rho}_{(\nu)}+\bar{P}_{(\nu)}}\delta \phi+3H{\dot{\bar{P}}_{(j)}\over \dot{\bar{\rho}}_{(j)}}(v_{(j)}+B)-H(v_{(j)}+B)\nonumber \\
&&-{\varphi \over a}-{\delta P_{(j)}\over a(\bar{\rho}_{(j)}+\bar{P}_{(j)})}-\dot{B},
\end{eqnarray}
is obtained. The evolution equation of the scalar field perturbation is
\begin{eqnarray}\label{p6}
&&\delta\ddot{\phi}+3H\delta\dot{\phi}+V''(\bar{\phi})\delta\phi+{k^2\over a^2}\delta \phi-(k^2\dot{E}+3\dot{\psi})\dot{\bar{\phi}}+{k^2\over a}B\dot{\bar{\phi}}-\dot{\bar{\phi}}\dot{\varphi}\nonumber \\
&&+2V'(\bar{\phi})\varphi+2\varphi \beta_{(\nu)}(\bar{\phi})(\bar{\rho}_{(\nu)}-3\bar{P}_{(\nu)})+\beta_{(\nu)}(\bar{\phi})(\delta \rho_{(\nu)}-3\delta P_{(\nu)})\nonumber \\
&&+\beta'_{(\nu)}(\bar{\phi})(\bar{\rho}_{(\nu)}-\bar{P}_{(\nu)})\delta \phi=0.
\end{eqnarray}
By considering  the Einstein equation, one derives
\begin{eqnarray}\label{p7}
&&3H(\dot{\psi}+\varphi H)+{k^2\over a^2}\left(\psi +H(a^2\dot{E}-a B)\right)=\nonumber \\
&&-{1\over 2M_P^2}\left(\sum_{j}\delta\rho_{(j)}-\varphi \dot{\bar{\phi}}^2+\delta\dot{\phi}\dot{\bar{\phi}}+
V'(\bar{\phi})\delta\phi \right)
\end{eqnarray}
from the $0-0$ component, and
\begin{equation}\label{p8}
\dot{\psi}+\varphi H=-{1\over 2M_P^2}\left(\sum_j a(v_{(j)}+B)(\bar{\rho}_{(j)}+\bar{P}_{(j)})-\dot{\bar{\phi}}\delta \phi\right)
\end{equation}
from the $0-i$ components, and
\begin{eqnarray}\label{p9}
\ddot{\psi}+3H\dot{\psi}+H\dot{\varphi}+(3H^2+2\dot{H})\varphi={1\over 2M_P^2}\left(\sum_j \delta P_{(j)}-\varphi \dot{\bar{\phi}}^2+\delta\dot{\phi}\dot{\bar{\phi}}-V'(\bar{\phi})\delta \phi\right)
\end{eqnarray}
by taking the trace of the $i-j$ components. The trace-free part of $i-j$ gives
\begin{equation}\label{p10}
\dot{\sigma_s}+H\sigma_s-\varphi +\psi=0,
\end{equation}
in which $\sigma_s=a^2\dot{E}-aB$ is the scalar shear.

In the following, we choose the flat gauge $\psi=E=0$. We assume that the Universe is constituted of the cold pressureless matter with $w_c=0$ (baryonic+dark matter), the radiation with $w_r={1\over 3}$, the nonrelativistic massive neutrino with $w_{(\nu)}=0$, and the quintessence. Only the interaction between the scalar field and the massive neutrinos is taken into account. For the background we have
\begin{eqnarray}\label{p11}
\dot{\bar{\rho}}_{(r)}+4H\bar{\rho}_{(r)}&=&0\nonumber \\
\dot{\bar{\rho}}_{(c)}+3H\bar{\rho}_{(c)}&=&0\nonumber \\
\dot{\bar{\rho}}_{(\nu)}+3H\bar{\rho}_{(\nu)}&=&\beta_{(\nu)}(\bar{\phi})\bar{\rho}_{(\nu)}\dot{\bar{\phi}}\nonumber\\
\ddot{\bar{\phi}}+3H\dot{\bar{\phi}}+V'(\bar{\phi})&=&-\beta_{(\nu)}(\bar{\phi})\bar{\rho}_{(\nu)}.
\end{eqnarray}
For the densities, we obtain
\begin{eqnarray}\label{p12}
&&\delta \dot{\rho}_r=-4H\delta \rho_{(r)}+{4k^2\over 3a}(\hat{v}_{(r)}-B)\bar{\rho}_{(r)}\nonumber\\
&&\delta \dot{\rho}_{(c)}=-3H\delta \rho_{(c)}+{k^2\over a}(\hat{v}_{(c)}-B)\bar{\rho}_{(c)}\nonumber\\
&&\delta \dot{\rho}_{(\nu)}=-3H\delta \rho_{(\nu)}+{k^2\over a}(\hat{v}_{(r)}-B)\bar{\rho}_{(\nu)}+\beta_{(\nu)}(\bar{\phi})\bar{\rho}_{(\nu)}\delta \dot{\phi}+\beta_{(\nu)}(\bar{\phi})\delta \rho_{(\nu)}\dot{\bar{\phi}}\nonumber \\
&&+\beta'_{(\nu)}(\phi)\bar{\rho}_{(\nu)}\dot{\bar{\phi}}\delta \phi,
\end{eqnarray}
and for the velocities, $\hat{v}_{(j)}=v_{(j)}+B$, we derive
\begin{eqnarray}\label{p13}
\dot{\hat{v_{(r)}}}&=&-{\varphi\over a}-{\delta \rho_{(r)}\over 4a\rho_{(r)}}\nonumber\\
\dot{\hat{v_{(c)}}}&=&-{\varphi\over a}-H\hat{v}_{(c)}\nonumber \\
\dot{\hat{v}}_{(\nu)}&=&-{\varphi\over a}-H\hat{v}_{(\nu)}-\beta_{(\nu)}(\bar{\phi}){\delta \phi \over a}.
\end{eqnarray}
The scalar field perturbation satisfies
\begin{eqnarray}\label{p14}
&&\delta \ddot{\phi}=-3H\delta \dot{\phi}-V''(\bar{\phi})\delta \phi-{k^2\over a^2}\delta \phi-{k^2\over a}\dot{\bar{\phi}}B-2V'(\bar{\phi})\varphi-2\varphi \beta_{(\nu)}(\bar{\phi})\bar{\rho}_{(\nu)}\nonumber \\
&&+{\dot{\bar{\phi}}\over 2HM_P^2}\left(\sum_j \delta P_j-\varphi \dot{\bar{\phi}}^2+\delta\dot{\phi}\dot{\bar{\phi}}-V'(\bar{\phi})\delta \phi\right)
-{3H^2+2\dot{H}\over H}\varphi \dot{\bar{\phi}}\nonumber \\
&&-\beta_{(\nu)}(\bar{\phi})\delta \rho_{(\nu)}-\beta'_{(\nu)}(\bar{\phi})(\bar{\rho}_{(\nu)}.
\end{eqnarray}
By using the components of the Einstein equation we get
\begin{equation}\label{p15}
\varphi=-{1\over 2HM_P^2}\left( -\dot{\bar{\phi}}\delta \phi +a\sum_j\hat{v}_{(j)}(\bar{\rho}_{(j)}+\bar{P}_{(j)})\right),
\end{equation}
and
\begin{eqnarray}\label{p16}
B&=&{3a\over 2k^2M_P^2}\left({1\over 3H}\left(\sum_j\delta \rho_{(j)}-\varphi \dot{\bar{\phi}}^2+\delta \dot{\phi} \dot{\bar{\phi}}+V'(\bar{\phi})\delta\phi\right)+\dot{\bar{\phi}}\delta\phi \right)\nonumber \\
&-&{3a^2\over 2k^2M_P^2}\sum_j\hat{v}_{(j)}(\bar{\rho}_{(j)}+\bar{P}_{(j)}).
\end{eqnarray}

In models in which the coupling of neutrinos and quintessence acts as a potential barrier, and forces the quintessence to trace the minimum of its effective potential, neutrino perturbations grow significantly in the nonrelativistic regime where the adiabaticity condition is used. In our model, the adiabatic condition does not hold, and instead, we use the slow roll condition. So we expect that the model is still stable against linear perturbation \cite{afsh}. Now let us numerically show this issue via the example (\ref{14.1}) introduced in the previous section.

 To numerically plot the perturbations, we also need to know the initial conditions for fluid velocities, energy density perturbations and perturbation of the scalar field.
At $\tau=tH_0=0$, we take
\begin{equation}\label{IC2}
\delta_{(\nu)}=\delta_c={3\over4 }\delta_{(r)}= 10^{-7},\,\,\delta \phi=10^{-7}\phi,\,\, \delta \dot{\phi}=10^{-7}\dot{\phi},\,\, \hat{v}_{(j)}=10^{-7}H_0^{-1},
\end{equation}
where $\delta_i={\delta_{\rho(i)}\over \rho_{(i)}}$ is the density contrast of the {\it{i}}th species.
A same initial 3-velocity $\hat{v}_{(i)}$  is assumed for all fluids. We have employed adiabatic initial conditions which imply  $\delta_c={3\over4 }\delta_r$ initially \cite{pert3}.

 The parameters and initial conditions are taken the same as in the previous section, i.e, $\{\alpha=15M_P^{-2},\,\,\,\beta=15M_P^{-2},\,\,\,\, V_0=0.691 \times 3H_0^2 M_P^2=2.74\times 10^{-47} GeV^4 \}$, and (\ref{IC1}).

\begin{figure}[H]
\centering\epsfig{file=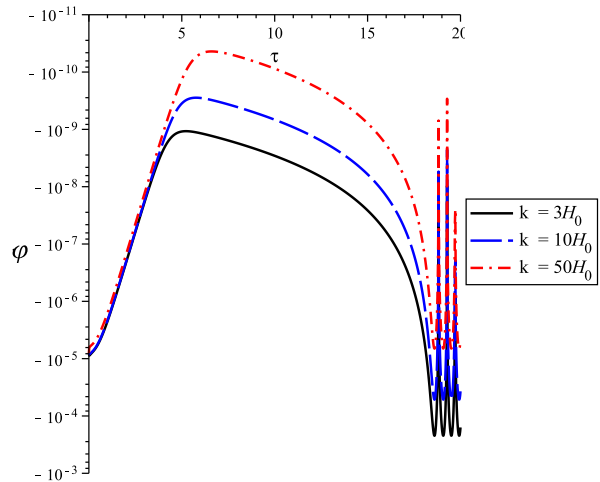,width=8cm,angle=0}
\caption{ Lapse function, $\varphi$, in terms of dimensionless time $\tau=tH_0$, with initial conditions (\ref{IC1}) and (\ref{IC2}) and $\{\alpha=15M_P^{-2},\,\,\,\beta=15M_P^{-2},\,\,\,\, V_0=0.691 \times 3H_0^2 M_P^2.\}$ } \label{fig6}
\end{figure}

\begin{figure}[H]
\centering\epsfig{file=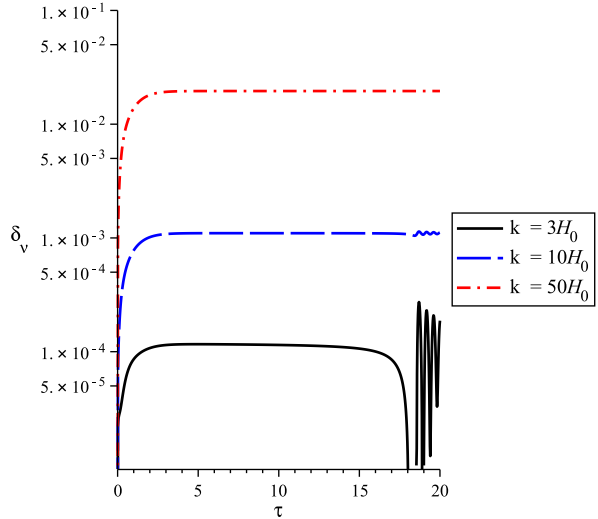,width=8cm,angle=0}
\caption{ Massive neutrinos density contrast in terms of dimensionless time $\tau=tH_0$ for three different wave numbers with initial conditions (\ref{IC1}), (\ref{IC2}) and $\{\alpha=15M_P^{-2},\,\,\,\beta=15M_P^{-2},\,\,\,\, V_0=0.691 \times 3H_0^2 M_P^2.\}$ } \label{fig7}
\end{figure}

In Fig.(\ref{fig6}) and  Fig.(\ref{fig7}), using Eqs. (\ref{p11})-(\ref{p16}),  we depict $\varphi$  and the massive neutrino density contrast, respectively

As it is evident from Fig.(\ref{fig7}), the neutrino perturbation does not grow critically. This is in contrast to the mass varying models of dark energy based on adiabaticity, where the linear perturbations in the nonrelativistic era grow and give rise to instability and formation of neutrino nuggets \cite{afsh}.

\section{Summary}
Inspired by the mass varying neutrino and symmetron models, we propose a new possible {\it{dynamical model}} of dark energy to describe the onset of the present cosmic acceleration. We assume that the quintessence is initially trapped in the minimum of its potential which has a $Z_2$ symmetry. In this era, both the kinetic and potential energies of the quintessence are negligible. This initial zero density is in agreement with the present astrophysical data which imply that despite the slower redshift of dark energy density with respect to the dark matter, they have the same order of magnitude today(pointed out in the coincidence problem).   After their relativistic era, the mass varying neutrinos become nonrelativistic, and the shape of the effective potential changes and the initial stable point becomes unstable. Contrary to the symmetron model, the effective potential and the potential have opposite slopes hence the quintessence climbs its potential while it rolls down the effective potential. This procedure provides enough energy to drive the cosmic acceleration via a slow roll evolution.

The quintessence-neutrino coupling modifies the evolution of the quintessence and consequently the dilution of dark energy. In the mass growing neutrino model, the scalar field dilutes like the dark matter during a significant period of its evolution, and therefore the coincidence problem may be alleviated \cite{Wet1}.
In our model, the beginning of quintessence evolution depends on the initial neutrino mass. It is only after the nonrelativistic epoch that the quintessence can commence its evolution from zero density to gain the same order of magnitude as the dark matter in later times.  In this way, one may relate the coincidence problem to the neutrino mass. The coincidence problem also depends on the other parameters of the model, especially those determining the dark energy density. To fix the parameters, we need to confront our model with observation data.

To illustrate how the model works, we used the example(\ref{14.1}) and chose the parameters, e.g. the initial neutrino mass, such that the derived present ratio densities are in agreement with the Planck 2015 data and the acceleration begins at $z\simeq 0.6$.  In a time of the order of the Hubble time, the dark energy density is given by $V_0$, which plays the role of a cosmological constant. So we fixed it as the value of the present dark energy density.  In this period, the model behaves like $\Lambda$CDM, but in contrast to the $\Lambda$CDM model, we have a dynamical dark energy with an initial zero density. Also, unlike the $\Lambda$CDM model, the acceleration is not persistent and by dilution of massive neutrinos the quintessence roles back to its initial position and oscillates about that position. However, to construct our model, we have to fine-tune our parameters like $V_0$, according to the astrophysical data. Note that it is also possible to consider other potentials, like those with an unbounded upper bound such as $V=V_0 (e^{\alpha\phi^2}-1),\,\, \alpha>0$. In these cases, like the example (\ref{14.1}), the quintessence climbs its own potential after the symmetry breaking, but unlike (\ref{14.1}) the potential does not have a maximum. The potential reaches at $V=V_0 (e^{\alpha\phi_{present}^2}-1)$ in the present era, which during the slow roll evolution may be identified with the present dark energy density. Again, by dilution of neutrinos, the quintessence will come back to its initial position and will oscillate about it via an underdamped oscillation.

In our scenario, as we do not employ the adiabaticity condition used in some of the previous models of neutrino dark energy, we do not encounter the instabilities that arise in those models. This issue was discussed and illustrated via numerical methods by using the example (\ref{14.1}).

\end{document}